\documentclass[twocolumn]{revtex4}
\usepackage{graphicx}
\usepackage{bm}
\usepackage{graphicx}
\usepackage{amsmath}
\usepackage{mathrsfs}
\usepackage{ulem}
\usepackage{color}
\usepackage[colorlinks, linkcolor=red,anchorcolor=green,citecolor=blue]{hyperref}
\usepackage{dcolumn,booktabs,bm}
\usepackage{amsfonts,amssymb,stmaryrd,latexsym,amsmath}
\usepackage{textcomp}
\usepackage{multirow}
\usepackage{diagbox}
\usepackage{array}

\begin{document}

\title{Charmonium Triangular Flow in High Energy Nuclear Collisions}

\author{Jiaxing Zhao$^a$}
\author{Baoyi Chen$^b$}
\author{Pengfei Zhuang$^a$}
\affiliation{$^a$Physics Department, Tsinghua University, Beijing 100084, China}
\affiliation{$^b$Department of Physics, Tianjin University, Tianjin 300350, China}

\begin{abstract}
We calculate, for the first time, $J/\psi$ triangular flow in high energy nuclear collisions. The charmonium motion in medium is controlled by a transport equation with loss and gain terms, and the evolution of the hot medium is governed by a single-shot hydrodynamic equation with a triangularly deformed initial condition. In comparison with the elliptic flow $v_2$, the triangular flow $v_3$ comes almost completely from the charmonium regeneration in the quark gluon plasma and therefore is more sensitive to the medium properties. 
\end{abstract}

\maketitle

\section{Introduction}
\label{s1}
Quarks and gluons can be deconfined at high temperature and form a new phase of strongly interacting matter, called quark-gluon plasma (QGP)~\cite{Aoki:2006we,Bazavov:2011nk}. The only way to generate the new state of matter in laboratory is through high energy nuclear collisions~\cite{STAR:2005gfr,PHENIX:2004vcz,ALICE:2010suc}. From the study on the collisions at the Relativistic Heavy Ion Collider (RHIC) and Large Hadron Collider (LHC), one of the characteristic observables to describe the collective properties of the QGP medium is the anisotropic flow of the final state hadrons~\cite{Ollitrault:1992bk}, which is defined as the coefficients $v_n$ of a Fourier decomposition of the hadron distribution~\cite{Voloshin:1994mz}, 
\begin{eqnarray}
\label{flow}
{dN\over d^2{\bm p}_T dy}={1\over \pi dp_T^2dy}\left[1+2\sum_{n=1}^\infty v_n \cos\left(n\left(\phi-\Psi_n\right)\right)\right],
\end{eqnarray}
where $y, {\bm p}_T$ and $\phi$ are respectively the hadron longitudinal rapidity, transverse momentum and azimuth angle, and $\Psi_n$ is the $n$-th harmonic symmetry plane angle. The momentum anisotropy of the produced hadrons in the final state originates from the spatial anisotropy of the collision overlap region in the initial state~\cite{Ollitrault:1992bk,Voloshin:1994mz,Qiu:2011iv,Luzum:2008cw}. For instance, the elliptic flow $v_2$ is a response to the ellipsoidal shape of the overlap region in non-central nucleus-nucleus collisions, and the triangular flow $v_3$ is mostly from the initial fluctuation of the energy density in the overlap region~\cite{Alver:2010dn,Alver:2010gr,Teaney:2010vd}. On the other hand, the collective flow develops in the medium and is therefore sensitive to the QGP properties, like the equation of state and shear and bulk viscosity~\cite{McDonald:2016vlt,Shen:2015qta}. 

Charmonia, the bound state of $c$ and $\bar c$ quarks, have long been considered as a sensitive signature of the QGP formation in nuclear collisions~\cite{Matsui:1986dk}. Different from light hadrons which are all produced through the decay of the medium at the phase transition boundary~\cite{Andronic:2006ky}, there are two sources for charmonium production: one is the initial production through hard processes which is then largely suppressed in the medium~\cite{Gerschel:1988wn,NA50:1996lag}, and the other is the regeneration in the medium through coalescence mechanism~\cite{Thews:2000rj,Grandchamp:2002wp,Andronic:2003zv,Yan:2006ve,Zhao:2017yan}. As colorless particles, the initially created quarkonia interact weakly with the QGP and their collective flow is very small. On the contrary, colored charmed quarks interact strongly with the medium and the regenerated charmonia can carry sizeable flow~\cite{Zhou:2014kka,Chen:2018kfo}. The competition between the suppression and regeneration can explain well~\cite{Zhao:2010nk,Zhao:2011cv,Liu:2009wza,Zhou:2014kka,Du:2015wha,Chen:2018kfo} the experimentally measured $J/\psi$ nuclear modification factor $R_{AA}$ and elliptic flow $v_2$ at RHIC and LHC energies, especially in the low momentum region. 

Recently, the inclusive $J/\psi\ v_3$ at forward rapidity and $v_2$ at middle and forward rapidity are measured in Pb-Pb collisions at colliding energy $\sqrt{s_{NN}}$= 5.02 TeV by ALICE Collaboration at LHC~\cite{ALICE:2020pvw,ALICE:2018bdo}. The first measurement of the charmonium triangular flow provides us an opportunity to probe the initial energy density fluctuation of the medium via heavy flavors and to understand the interaction mechanism of heavy quarks with the medium. In this paper we study the $J/\psi$ triangular flow in the frame of a transport approach. 

\section{Medium evolution}
\label{s2}
We first discuss the initial energy density of the medium. Since the medium is dominated by light partons, the feedback from charmonia to medium can be safely neglected, and the medium can be considered as a background for the charmonium motion. In this case we can take a well established hydrodynamic model for light hadron production to describe the evolution of the background. The initial energy profile of the medium without fluctuations can be factorized as~\cite{Schenke:2010nt}
\begin{eqnarray}
\label{epsilon0}
\epsilon({\bm x},\tau_0|{\bm b})=\epsilon_0 f(\eta)W({\bm x}_T|{\bm b})/W(0|0),
\end{eqnarray}
where $\epsilon_0$ is the maximum energy density at the center of the fireball created in central collisions, which is related to the maximum temperature $T_0$ via the equation of state and can be fixed by matching to the experimentally measured charged hadron multiplicity~\cite{Schenke:2010nt}. In our numerical calculation $T_0$ is taken to be $510$ MeV for Pb-Pb collisions at $\sqrt{s_{NN}}$=5.02\ TeV. The longitudinal and transverse space dependence is characterized by the distributions $f(\eta)$ and $W({\bm x}_T|{\bm b})$, where $\eta=1/2\ln[(t+z)/(t-z)]$ is the space-time rapidity and we take $f(\eta)$ as the one in Ref.~\cite{Schenke:2010nt}. The initial state of matter in the transverse plane is normally described by two components~\cite{Kharzeev:2000ph,Bozek:2010wt,Schenke:2010nt}, the number of wounded nucleons $n_\text{w}$ and the number of binary nucleon-nucleon collisions $n_\text{b}$, which respectively control the initial soft and hard processes. While how the deposited energy density or entropy density precisely scales with the two numbers is not clear from the first principle, we know that the soft process is the main source of the initial energy deposition at SPS energy, and the contribution from the hard process increases at RHIC and LHC energy. In general, the spatial dependence of the initial energy distribution in the transverse plane can be parameterized as~\cite{Schenke:2010nt},
\begin{eqnarray}
\label{wx}
W({\bm x}_T|{\bm b})=(1-\gamma)n_{\text{w}}({\bm x}_T|{\bm b})+\gamma n_{\text{b}}({\bm x}_T|{\bm b}),
\end{eqnarray}
where ${\bm x}_T$ is the transverse coordinate of the colliding nucleon, ${\bm b}$ the impact parameter of the nuclear collisions, and the parameter $\gamma$ the fraction of binary collisions. At LHC energy, one takes $\gamma=0.16$ for Pb-Pb collisions at $\sqrt{s_{NN}}$=5.02\ TeV~\cite{Bozek:2010wt}. The numbers of wounded nucleons and binary collisions are controlled by the nuclear geometry,  
\begin{eqnarray}
\label{nwnb}
n_{\text{w}}({\bm x}_T|{\bm b}) &=& T_A({\bm x}_T-{\bm b}/2)\left(1- e^{-\sigma_{\text{inel}} T_B({\bm x}_T+{\bm b}/2)}\right)\nonumber\\
&+& T_B({\bm x}_T+{\bm b}/2)\left(1- e^{-\sigma_{\text{inel}} T_A({\bm x}_T-{\bm b}/2)} \right), \nonumber\\
n_{\text{b}}({\bm x}_T|{\bm b}) &=& \sigma_{\text{inel}}T_A({\bm x}_T+{\bm b}/2)T_B({\bm x}_T-{\bm b}/2),
\end{eqnarray}
where $T_A$ and $T_B$ are the thickness functions~\cite{Miller:2007ri}, and $\sigma_{\text{inel}}$ is the inelastic scattering cross-section between nucleons. For Pb-Pb collisions at $\sqrt{s_{NN}}=5.02$\ TeV one takes $\sigma_{\text{inel}}=68\ \text{mb}$~\cite{ALICE:2012fjm}. 

In relativistic heavy ion collisions, the initial condition of the medium evolution fluctuates event by event, and the fluctuations can be simulated through the Monte Carlo Glauber model (MC-Glauber)~\cite{Miller:2007ri} and Monte Carlo fKLN model (MC-KLN)~\cite{Kharzeev:2001gp,Drescher:2006ca}. The QCD-based IP-Glasma model can also describe well the initial stage of the medium~\cite{Schenke:2012wb}. With these initial conditions including event-by-event fluctuations, the anisotropic flows for charged hadrons measured at RHIC and LHC are explained very well in the frame of relativistic viscous hydrodynamics followed by a transport cascade in hadronic phase~\cite{McDonald:2016vlt,Shen:2015qta}. 

\begin{figure}[htb]
{ $$\includegraphics[width=0.45\textwidth] {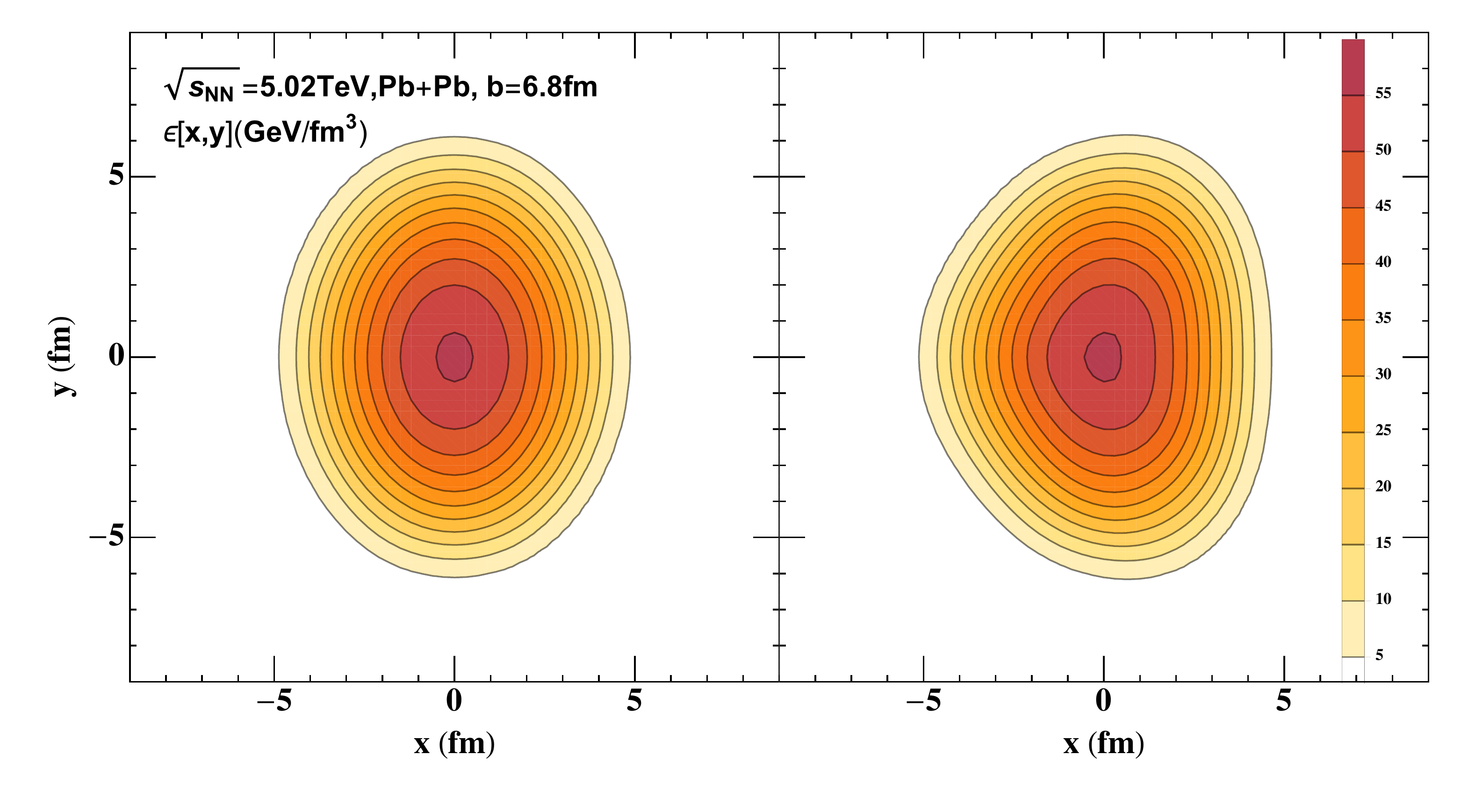}$$
\caption{The initial energy density $\epsilon(x,y)$ in the transverse plane for Pb-Pb collisions at $\sqrt{s_{NN}}=$5.02 TeV with impact parameters $b=6.8$ fm. The left and right panels are the profiles without and with triangular deformation. }
\label{fig1}}
\end{figure}

In principle, aiming to describe the $J/\psi$ triangular flow $v_3$, one needs to evolve the event-by-event hydrodynamics with fluctuating initial conditions~\cite{McDonald:2016vlt,Qiu:2011iv}. However, to simplify the calculations, one can take, as an approximation, a single-shot hydrodynamics where the initial condition is given by an averaged, smooth but deformed profile~\cite{Qiu:2011iv,Alver:2010dn,Alver:2010gr}. We follow the way shown in Ref.~\cite{Alver:2010dn} to add a deformation factor in the energy density through the transformation 
\begin{eqnarray}
\label{epsilon0+}
\epsilon({\bm x},\tau_0|{\bm b})\to \epsilon (\tilde{\bm x},\tau_0|{\bm b})
\end{eqnarray}
with the deformed coordinate $\tilde{\bm x}=(x_T\sqrt{1+\varepsilon_3\cos[3(\phi_s-\Psi_3)]},\phi_s,\eta)$, where $x_T=\sqrt{x^2+y^2}$ and $\phi_s=\arctan(|y|/|x|)$ are the transverse radius and azimuth angle, $\Psi_3$ is the reference angle which can be the reaction plane angle in smooth hydrodynamics (we take $\Psi_3=0$ in the following  calculation). The magnitude of the deformation $\varepsilon_3$, which is called triangularity and depends on the collision centrality, can be given by the MC-Glauber or MC-KLN model~\cite{Alver:2010dn,Qiu:2011iv}. We take $\varepsilon_3$= 0.08, 0.1 and 0.2 for Pb-Pb collisions at $\sqrt{s_{NN}}$=5.02\ TeV with impact parameters $b=$ 3.2, 6.8 and 9.6 fm, corresponding to the centrality bin 0-10\%, 10-30\% and 30-50\%~\cite{Alver:2010dn,Qiu:2011iv}.
The initial energy density profile with and without triangular deformation is shown in Fig.~\ref{fig1}. 

The quark and gluon medium produced in high energy nuclear collisions is a strongly coupled system and therefore can be locally thermalized fast. With the fluctuating energy density (\ref{epsilon0+}) as the initial condition at time $\tau_0=0.6$ fm/c~\cite{Schenke:2010nt}, the evolution of the medium can be described by hydrodynamic equations. In this paper, we employ a (2+1)-dimensional hydrodynamic model, the MUSIC package~\cite{Schenke:2010nt,McDonald:2016vlt}, to characterize the space and time dependence of the temperature and velocity of the hot medium. To close the hydrodynamic equations, the equation of state for both QGP and hadron phases is needed. We will adopt "s95p-v1" which matches Lattice QCD data at high temperature and the hadron resonance gas at low temperature~\cite{Huovinen:2009yb}. The two phases are connected with a smooth crossover. The critical temperature $T_c$ between QGP and hadron gas is taken to be 170 MeV. We choose an effective shear viscosity $\eta/s=0.08$~\cite{Bernhard:2016tnd} and a zero bulk viscosity. 

\section{Charmonium transport}
\label{s3}
We now focus on the charmonium motion in hot medium. Since quarkonia are colorless and very heavy, they are unlikely thermalized with the medium. Therefore, their phase space distribution $f_\psi({\bm p},{\bm x},\tau|{\bm b})$ for $\psi=J/\psi, \chi_c, \psi'$ should be governed by a transport equation including both initial production 
as well as regeneration. The distribution satisfies the relativistic Boltzmann equation~\cite{Zhao:2020jqu},
\begin{eqnarray}
\label{transport}
&&\left[ \cosh(y-\eta)\partial_\tau + {\sinh(y-\eta)\over \tau}\partial_\eta+{\bm v}_T\cdot \nabla_T \right] f_\psi\nonumber\\
= && -\alpha f_\psi +\beta,
\end{eqnarray}
where $y=1/2\ln[(E+p_z)/( E-p_z)]$ is the charmonium momentum rapidity, and ${\bm v}_T={\bm p}_T/E_T$ is the charmonium transverse velocity with transverse energy $E_T=\sqrt{m_{\psi}^2+{\bm p}_T^2}$. The second and third terms on the left hand side arise from the free streaming of $\psi$ which leads to the leakage effect in the longitudinal and transverse direction. The anomalous suppression and regeneration mechanisms in QGP medium are reflected in the loss term $\alpha$ and gain term $\beta$. 

Charmonia in hot QGP medium suffers Debye screening~\cite{Matsui:1986dk}. With increasing temperature, the interaction distance between a pair of heavy quarks becomes shorter and shorter. When it is less than the charmonium size, the charmonium is dissociated. Considering different sizes for the ground and excited states of $c\bar c$, the dissociation temperatures are $T_d\sim (2.3, 1.2, 1.1)T_c$ for $J/\psi$, $\chi_c$ and $\psi'$. This is the so-called sequential suppression~\cite{Satz:2005hx}. The above analyses based on the Debye screening effect is typically based on the assumption of a constant temperature in connection with a sharp transition of the inelastic charmonium widths from zero below $T_d$ to infinity above $T_d$. However, the volume of the produced fireball in relativistic heavy ion collisions is relatively small and expands rapidly, implying rather fast temperature changes and short fireball lifetimes. In this case, the conclusion from the static Debye screening effect may deviate from the real system, and it becomes essential to include the concrete interactions between partons and charmonia, leading to sizable inelastic reaction rates comparable to the fireball expansion or cooling rate. In particular, the charmonia can be destroyed below $T_d$ and survivable above $T_d$. The Debye screening is still operative, by controlling the binding energy which in turn determines the phase space and thus the width of the dynamic dissociation reactions~\cite{Satz:2005hx,Chen:2017jje}. An important such process in the QGP is the gluon dissociation process~\cite{Peskin:1979va,Bhanot:1979vb} $g\psi \to c \bar c$, an analogy to the photon dissociation process of electromagnetic bound states. For the ground state $J/\psi$, the gluon dissociation cross-section $\sigma_{g\psi}^{c\bar c}$ in vacuum is derived using the operator-production-expansion (OPE) method~\cite{Peskin:1979va,Bhanot:1979vb}. For the excited states $\chi_c$ and $\psi'$, the cross sections can be obtained through their geometric relation to the ground state~\cite{Chen:2018kfo}. Taking only the gluon dissociation as the loss term and its inverse process $c\bar c\to g\psi$ as the gain term (both are $2\to 2$ processes), $\alpha$ and $\beta$ can be explicitly expressed as~\cite{Zhao:2020jqu}
\begin{eqnarray}
\label{alphabeta}
\alpha({\bm p},{\bm x},\tau|{\bf b}) &=& {1\over 2E_T}\int{d^3{\bm p}_g \over (2\pi)^3 2E_g}W_{g\psi}^{c\bar c}(s)f_g({\bm p}_g,{\bm x},\tau)\nonumber\\
&&\times\Theta(T({\bm x},\tau|{\bm b})-T_c),\nonumber\\
\beta({\bm p},{\bm x},\tau|{\bm b}) &=& {1\over 2E_T}\int {d^3{\bm p}_g \over (2\pi)^3 2E_g}{d^3{\bm p}_c \over(2\pi)^3 2E_c}{d^3{\bm p}_{\bar c} \over(2\pi)^3 2E_{\bar c}}\nonumber\\
&&\times W_{c\bar c}^{g\psi}(s)f_c({\bm p}_c,{\bm x},\tau|{\bm b})f_{\bar c}({\bm p}_{\bar c},{\bm x},\tau|{\bm b})\nonumber\\
&&\times(2\pi)^4\delta^{(4)}(p+p_g-p_c-p_{\bar c})\nonumber\\
&&\times\Theta(T({\bm x},\tau|{\bm b})-T_c),
\end{eqnarray}
where $E_g, E_c$ and $E_{\bar c}$ are the gluon, charm quark and anti-charm quark energies, ${\bm p}_g, {\bm p}_c$ and ${\bm p}_{\bar c}$ are their momenta, and $s$ is the $g\psi$ interaction energy in their center-of-mass frame. $W_{g\psi}^{c\bar c}$ is the dissociation probability, and $W_{c\bar c}^{g\psi}$ the regeneration probability which can be obtained from $W_{g\psi}^{c\bar c}$ through considering detailed  balance between the two processes~\cite{Yan:2006ve,Liu:2009wza}. The step function $\Theta$ is used to guarantee the calculation in QGP phase above the critical temperature $T_c$, and the local temperature of the medium comes from the hydrodynamic calculation discussed in Section \ref{s2}.
Since gluons are constituents of the medium, their distribution $f_g$ is the Bose-Einstein distribution. Again the local temperature and velocity of the medium in the distribution are from the hydrodynamics. Charm quarks are not constituents of the medium, their motion in the medium should in principle be controlled by a transport equation, and the distribution is in between two limits: the pQCD limit without interaction with the medium and the equilibrium limit with strong interaction with the medium. The experimentally measured large charmed meson flow indicates that charm quarks with small transverse momentum might be thermalized with the medium~\cite{STAR:2017kkh,ALICE:2020pvw}. As a first approximation, we take a thermally equilibrated distribution for charm (anti-charm) quarks  $f_c=\rho_c N_c/(e^{p_c^\mu u_\mu/T}+1)$, where $N_c$ is the normalization factor, and the density in coordinate space $\rho_c$ is controlled by the charm conservation equation $\partial_\mu(\rho_cu^\mu)=0$. The initial density is governed by the nuclear geometry of the colliding system,  
\begin{equation}
\label{charm}
\rho_c({\bm x},\tau_0|{\bm b})={T_A(\tilde {\bm x}_T+{\bf b}/2)T_B(\tilde {\bm x}_T-{\bm b}/2)\cosh \eta \over \tau_0}{d\sigma_{pp}^{c\bar c}\over d\eta},
\end{equation}
where $d\sigma^{c\bar{c}}_{pp}/d\eta$ is the rapidity distribution of charm quark production cross section in p-p collisions. We take 0.56 mb $< d\sigma^{c\bar{c}}_{pp}/d\eta<$ 0.78 mb in the forward rapidity of Pb-Pb collisions at $\sqrt{s_{NN}}=5.02$ TeV~\cite{LHCb:2016ikn, ALICE:2021dhb}. Since charm quarks are produced through initial hard scattering processes controlled by the number of binary collisions $n_{\text{b}}$, their initial density should also be triangularly deformed. This is introduced via the change from ${\bm x}$ to $\tilde{\bm x}$ in $\rho_c$.  

Besides the hot nuclear matter effect which affects charmonium motion through the above discussed suppression and regeneration terms ($\alpha$ and $\beta$), there is also the cold nuclear matter effect which changes the initial condition of the transport equation (\ref{transport}). If the cold effect is completely omitted, the initial condition is just a superposition of the charmonium distribution in p-p collisions which can be parameterized as~\cite{Zhou:2014kka,Chen:2018kfo,Zhao:2017yan},
\begin{eqnarray}
\label{pp}
&&{d^2\sigma^{J/\psi}_{pp}\over 2\pi p_Tdp_Tdy}\nonumber\\
= &&{n-1\over n-2}{1\over \pi\langle p_T^2\rangle_{pp}}\left(1+{1\over n-2}{p_T^2\over \langle p_T^2\rangle_{pp}}\right)^{-n}{d\sigma^{J/\psi}_{pp}\over dy},
\end{eqnarray}
where the mean transverse momentum square $\langle p_T^2\rangle_{pp}=$ 12.5 (GeV/c)$^2$, the parameter $n=3.2$, and the rapidity distribution $d\sigma^{J/\psi}_{pp}/ dy=$ 3.25 $\mu$b in forward rapidity $2.5<|y|<4$ of Pb-Pb collisions at $\sqrt{s_{NN}}$=5.02 TeV are extracted from the experimental data~\cite{ALICE:2012vup,ALICE:2012vpz}. The cold nuclear matter effect on charmonium production includes mainly nuclear shadowing~\cite{Mueller:1985wy}, Cronin effect~\cite{Cronin:1974zm,Hufner:1988wz} and nuclear absorption~\cite{Gerschel:1988wn}. For heavy ion collisions at LHC energy, the collision time is very short, the nuclear absorption can be safely neglected. Before two gluons fuse into a charmonium, they acquire additional transverse momentum via multi-scattering with the surrounding nucleons, and this extra momentum would be inherited by the produced charmonium, this is the so-called Cronin effect. Therefore, when we do the superposition of the p-p distribution (\ref{pp}), the averaged transverse momentum square $\langle p_T^2\rangle_{pp}$ should be replaced by 
 \begin{equation}
 \label{cronin}
\langle p_T^2\rangle_{pp}+a_{gN}l,
 \end{equation}
where the Cronin parameter $a_{gN}$ is the averaged charmonium transverse momentum square obtained from the gluon scattering with a unit of length of nucleons, and $l$ is the mean trajectory length of the two gluons in the projectile and target nuclei before $c\bar c$ formation. We take $a_{gN} = 0.15\text{GeV}^2/\text{fm}$ for Pb-Pb collisions at $\sqrt{s_{NN}}$=5.02 TeV. In practice, we take a Gaussian smearing~\cite{Huefner:2002tt,Zhao:2010nk,Zhou:2014kka} for the modified transverse momentum distribution. The shadowing effect comes from the parton collective motion in a nucleus~\cite{Norton:2003cb} and can be parameterized as a modification of the parton distribution function $f_i(x,\mu_F)$ in a free nucleon by a factor $R_i= \bar f_i/(Af_i)$, where $\bar f_i(x,\mu_F)$ is the distribution for parton $i$ in a nucleus, and the factorization scale is taken as $\mu_F=\sqrt{m_\psi^2+p_T^2}$. The modification factor can be taken from the EPS09 package~\cite{Helenius:2012wd}. Including both the shadowing and Cronin effects, the initial charmonium distribution can be written as~\cite{Zhou:2014kka}
 \begin{eqnarray}
 \label{initial}
&& f_\psi ({\bm p}, {\bm x}, \tau_0|{\bm b})\nonumber\\
= && {(2\pi)^3 \over E_T \tau_0}\int dz_A dz_B \rho_A(\tilde {\bm x}_T+{{\bm b}/2},z_A)\rho_B(\tilde {\bm x}_T-{{\bm b}/2},z_B)\nonumber\\
&&\times\mathcal{R}_g(x_1,\mu_F, \tilde{\bm x}_T+{\bm b}/2)\mathcal{R}_g(x_2,\mu_F, \tilde{\bm x}_T-{\bm b}/2)\nonumber\\
&&\times f_{pp}^{\psi}(\tilde{\bm x}_T,{\bm p},z_A,z_B|{\bm b}),
 \end{eqnarray}
where $\rho_A$ and $\rho_B$ are the nucleon distributions in the two colliding nuclei, the charmonium momentum distribution $f_{pp}^{\psi}$ is with the Cronin-effect modified averaged transverse momentum square (\ref{cronin}). The local shadowing effect is embedded in the factor $\mathcal R_g$. Again, the triangular deformation is considered in the initial distributions via the replacement of
${\bm x}_T$ by the deformed coordinate $\tilde{\bm x}_T$.  

\section{Numerical result}
\label{s4}
The transport equation (\ref{transport}) with the initial distribution (\ref{initial}) can be solved analytically~\cite{Yan:2006ve,Liu:2009wza}. To compare the result with the experimentally measured inclusive $J/\psi$ data, one should consider not only the above discussed direct production but also the feed-down from the excited charmonium states and the B-decay contribution. The feed-down branch ratio from $\chi_c$ and $\psi'$ to $J/\psi$ are taken as 22\% and 61\%~\cite{ParticleDataGroup:2020ssz}, and the momentum dependence of the B-decay fraction in p-p collisions can be fitted as $f_{B}(p_T)=N_{pp}^{B\to J/\psi}/(N_{pp}^{\text{prompt}} + N_{pp}^{B\to J/\psi})=0.04 + 0.023p_T/(\text{GeV/c})$ which depends on rapidity and colliding energy weakly~\cite{Chen:2013wmr}. We take the same fraction in heavy ion collisions. For bottom quarks, their initial momentum distribution can be given by FONNL~\cite{FONLL} and the evolution in QGP can be simulated by the Langevin equation with a spatial diffusion coefficient $2\pi TD_s=2$~\cite{Zhao:2020jqu}. After the evolution bottom quarks will carry an anisotropic flow $v_n^{\text B} (n=2,3)$. With both prompt (direct + feed-down) and non-prompt (B-decay) contributions, the $J/\psi$ anisotropic flow $v_n$ in the final state can be expressed as,
 \begin{eqnarray}
v_n(p_T)=v_n^{\text{prompt}}(p_T)\left(1-f_B(p_T)\right)+v_n^{\text{B}}(p_T)f_B(p_T).
 \end{eqnarray}
 
\begin{figure}[htb]
	{ \includegraphics[width=0.38\textwidth] {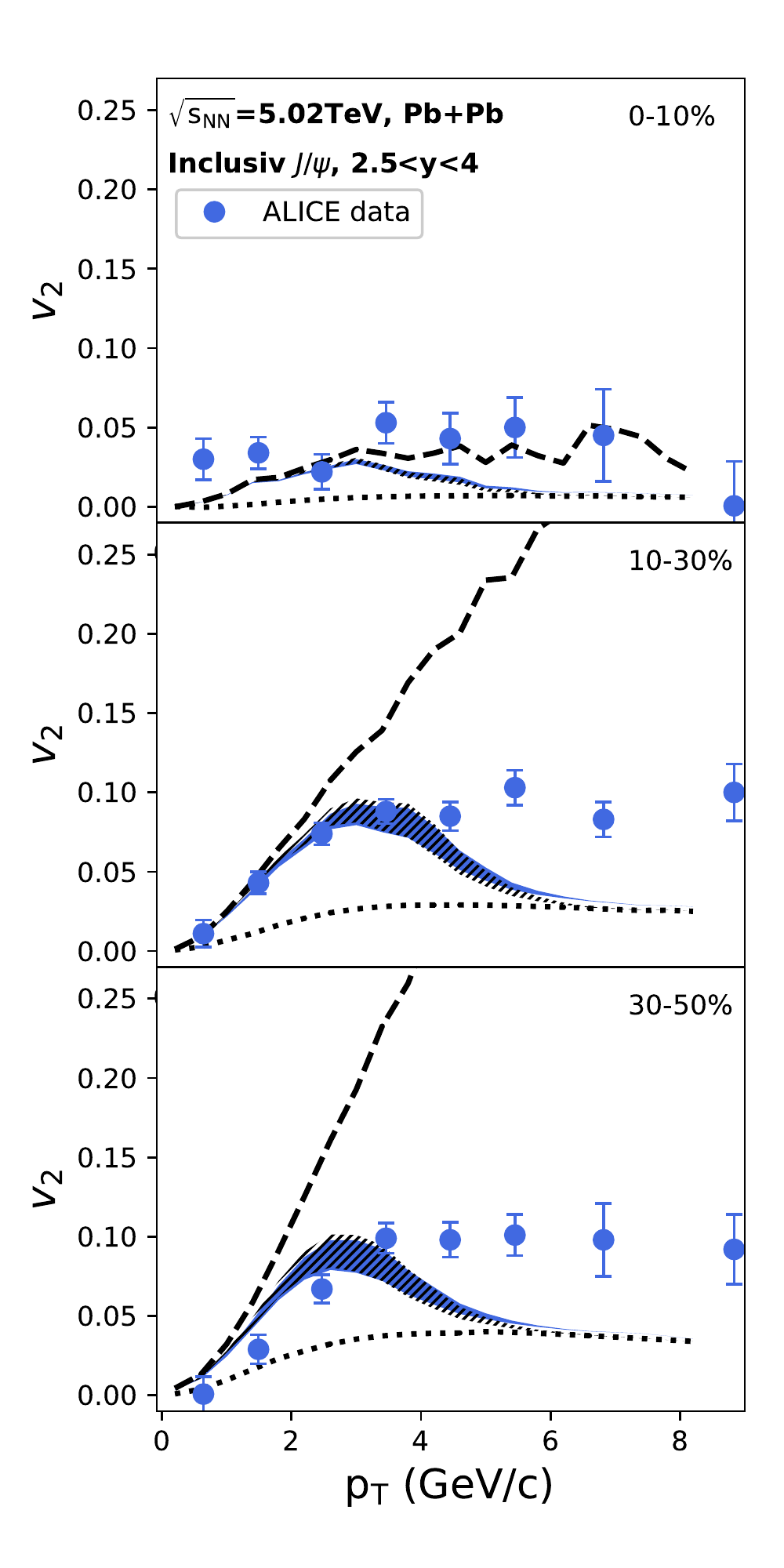}
		\caption{The $J/\psi$ elliptic flow $v_2$ as a function of transverse momentum $p_T$ in different centrality bins in Pb-Pb collisions at $\sqrt{s_{NN}}$= 5.02 TeV. Dotted and dashed lines are calculations with only initial production and regeneration, slash and dark bands are total results without and with B-decay contribution, and the lower and higher limits of the bands correspond to charm quark cross section $d\sigma^{c\bar c}_{pp}/d\eta=$ 0.56 and 0.78 mb. The experimental inclusive data are from ALICE Collaboration~\cite{ALICE:2020pvw}.}
		\label{fig2}}
\end{figure}
\begin{figure}[htb]
	{ \includegraphics[width=0.38\textwidth] {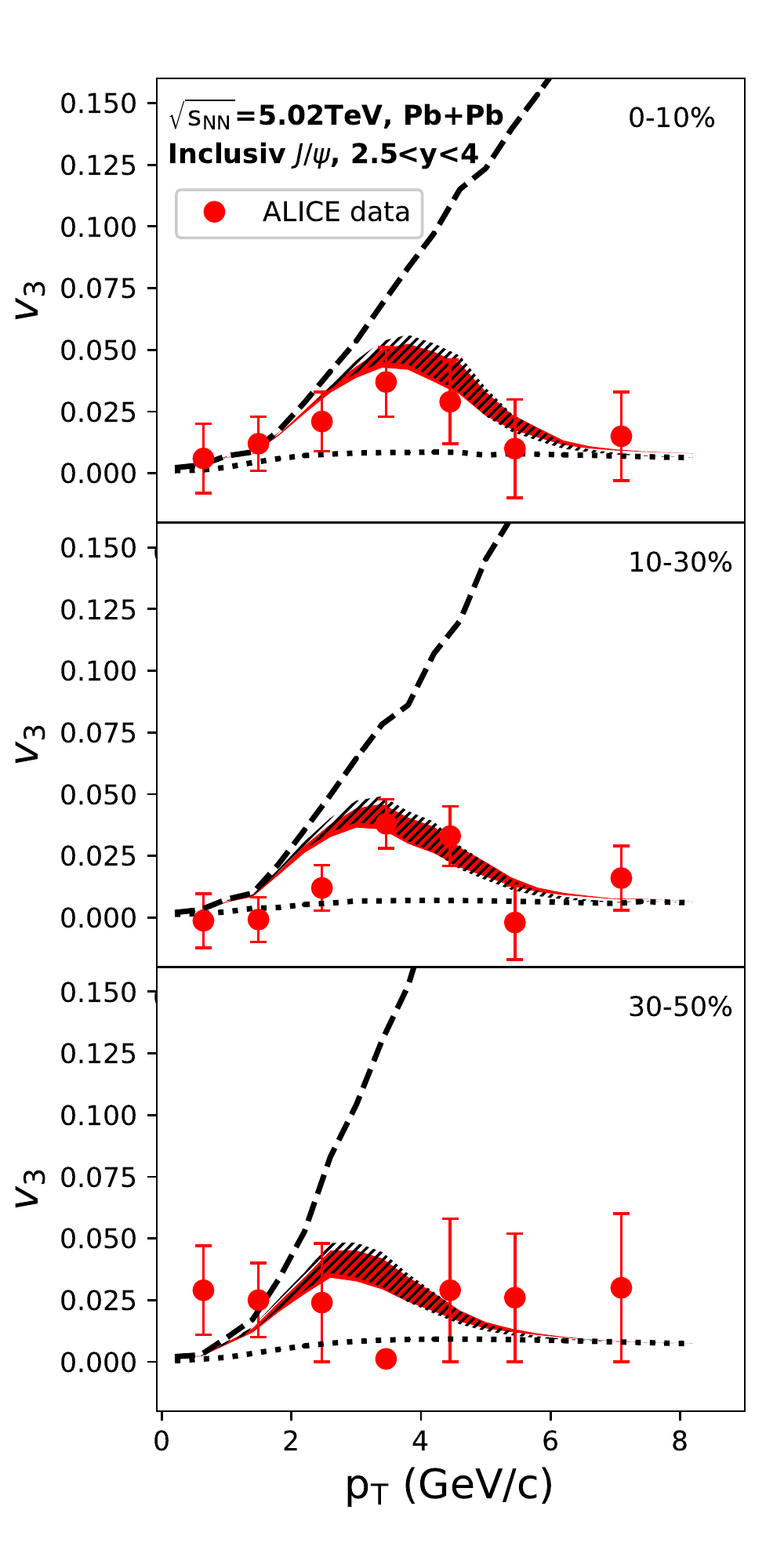}
		\caption{The $J/\psi$ triangular flow $v_3$ as a function of transverse momentum $p_T$ in different centrality bins in Pb-Pb collisions at $\sqrt{s_{NN}}$= 5.02 TeV. Dotted and dashed lines are calculations with only initial production and regeneration, slash and dark bands are total results without and with B-decay contribution, and the lower and higher limits of the bands correspond to charm quark cross section $d\sigma^{c\bar c}_{pp}/d\eta=$ 0.56 and 0.78 mb. The experimental inclusive data are from ALICE Collaboration~\cite{ALICE:2020pvw}.}
		\label{fig3}}
\end{figure}
The numerical calculations for $J/\psi\ v_2$ and $v_3$ and the comparison with inclusive $J/\psi$ data in Pb-Pb collisions at $\sqrt{s_{NN}}$= 5.02 TeV are shown in Figs.~\ref{fig2} and~\ref{fig3}. Dotted and dashed lines are the calculations with only initial production and regeneration, slash and dark bands are the total results for prompt and inclusive $J/\psi$s, and the higher and lower limits of the bands correspond to the charm quark cross section $d\sigma_{pp}^{c\bar c}/d\eta=0.78$ and $0.56$ mb. Considering the fact that, the initially produced charmonia via hard processes carry high momentum, and the regeneration via coalescence of thermalized charm quarks happens at low momentum, their contributions to charmonium production are mainly in the higher and lower momentum regions. Since bottom quark is too heavy, its decay contribution to $J/\psi$ yield is small, and its interaction with the medium is much weaker than a charm quarks which leads to a rather small $v_n^\text B$. That is the reason why the difference between the slash and dark bands is very small. Our calculated $v_2$ agrees reasonably well with the experimental data at $p_T\lesssim 3$ GeV/c, the sizeable $v_2$ at high momentum may come from charm quark interaction with the magnetic field created in non-central collisions~\cite{Fukushima:2015wck} and the fragmentation mechanism which dominates the high-momentum charmonium production~\cite{Bain:2016rrv,Kang:2017yde}.              

Different from the elliptic flow $v_2$ which is initiated from the anisotropic nuclear geometry and thus increases from central to peripheral collisions, the source of the triangular flow $v_3$ is the initial fluctuation of the medium which is not directly related to the collision centrality. This difference leads to the following three characteristics of the $J/\psi$ triangular flow: 1) $v_3$ is sensitive to the hot medium formation, but its dependence on centrality is weak provided the fireball is formed; 2) The initial production is before the medium formation and the following interaction between colorless charmonia and the medium is weak, the contribution from initial production to $v_3$ is extremely small and $v_3$ is almost fully from the regeneration; 3) High $p_T$ charmonia are not from the regeneration in medium and therefore not sensitive to the initial medium fluctuation, $v_3$ approaches to zero fast at high momentum. 

\section{Summary}
\label{summary}
Charmonium collective flow is often used as a probe of the hot medium properties in high energy nuclear collisions. We calculated $J/\psi$ elliptic flow $v_2$ and triangular flow $v_3$ in the frame of a transport approach for charmonium motion and an initially triangularly deformed hydrodynamics for the medium evolution. Different from $v_2$ which is originated from the anisotropic nuclear geometry in the beginning of the collisions, $v_3$ comes from the fluctuations of the initial condition of the hot medium and therefore is more sensitive to the medium properties. While there are different mechanisms for charmonium production in different momentum regions, like regeneration at low momentum and initial production and fragmentation at high momentum, the medium-fluctuation controlled $v_3$ exists only in low and intermediate momentum region and can be well described by the regeneration mechanism. 
\\

\noindent {\bf Acknowledgement}: The work is supported by the NSFC grant Nos. 11890712, 12047535, 12075129 and 12175165. 
 

\end{document}